\documentclass[aps,prd,preprint,groupedaddress,showpacs]{revtex4}
\usepackage{epsfig}
\usepackage{graphics}
\usepackage{slashed}

\begin{document}

\leftmargin -2cm
\def\choosen{\atopwithdelims..}

\preprint{DESY~11--115\hspace{12.0cm} ISSN 0418--9833}
\preprint{NSF--KITP--11--119\hspace{13.9cm}}

\title{Single jet and prompt-photon inclusive production with multi-Regge
kinematics: From Tevatron to LHC}

\author{\firstname{B.A. }\surname{Kniehl}}\thanks{On leave of absence from
II. Institut f\"ur Theoretische Physik, Universit\" at Hamburg,
Luruper Chaussee 149, 22761 Hamburg, Germany}

\email{kniehl@desy.de}
\affiliation{Kavli Institute for Theoretical Physics, Kohn Hall,
University of Santa Barbara, CA~93106, USA}

\author{\firstname{V.A. }\surname{Saleev}}
\email{saleev@ssu.samara.ru}
\affiliation{Samara State University, Ac.~Pavlov, 1, 443011, Samara, Russia}
\affiliation{S.P.~Korolyov Samara State Aerospace University, Moscow highway,
34, 443086, Samara, Russia}

\author{\firstname{A.V. }\surname{Shipilova}}
\email{alexshipilova@ssu.samara.ru}
\affiliation{Samara State University, Ac.~Pavlov, 1, 443011, Samara, Russia}

\author{\firstname{E.V. }\surname{Yatsenko}}
\email{elena.yatsenko@desy.de}
\affiliation{{II.} Institut f\"ur Theoretische Physik, Universit\" at Hamburg,
Luruper Chaussee 149, 22761 Hamburg, Germany}

\begin{abstract}
We study single jet and prompt-photon inclusive hadroproduction with
multi-Regge kinematics invoking the hypothesis of parton Reggeization in
$t$-channel exchanges at high energy.
In this approach, the leading contributions are due to the fusion of two
Reggeized gluons into a Yang-Mills gluon and the annihilation of a Reggeized
quark-antiquark pair into a photon, respectively.
Adopting the Kimber-Martin-Ryskin and Bl\"umlein prescriptions to derive
unintegrated gluon and quark distribution functions of the proton from their
collinear counterparts, for which we use the Martin-Roberts-Stirling-Thorne
set, we evaluate cross section distributions in transverse momentum ($p_T$)
and rapidity.
Without adjusting any free parameters, we find good agreement with measurements
by the CDF and D0 Collaborations at the Tevatron and by the ATLAS Collaboration
at the LHC in the region $2p_T/\sqrt S\alt 0.1$, where $\sqrt S$ is the
hadronic c.m.\ energy.
\end{abstract}


\maketitle

\section{Introduction}
\label{sec:one}

The study of jet and prompt-photon inclusive production at high-energy
colliders, such as the Fermilab Tevatron and the CERN LHC, is of great interest
because it allows us to test perturbative quantum chromodynamics (QCD) and to
extract information on the parton distribution functions (PDFs) of the proton.
The presence of a jet or a photon
with large transverse momentum, $p_T\gg \Lambda_\mathrm{QCD}$, with
$\Lambda_\mathrm{QCD}$ being the asymptotic scale parameter, guarantees that the
strong coupling constant remains small in the processes discussed here,
i.e.\ typically $\alpha_s(p_T)\alt 0.1$.

The total collision energies, $\sqrt{S}=1.8$~TeV and 1.96~TeV in Tevatron
runs~I and II, respectively, and $\sqrt{S}=7$~TeV or 14~TeV at the LHC,
sufficiently exceed the characteristic scale $\mu$ of the relevant hard
processes, which is of order of $p_T$, {\it i.e.}\ we have
$\Lambda_\mathrm{QCD}\ll\mu\ll\sqrt{S}$.
In this high-energy regime, the contribution of partonic subprocesses involving
$t$-channel parton (gluon or quark) exchanges to the production
cross section can become dominant.
Then the transverse momenta of the incoming partons and their off-shell
properties can no longer be neglected, and we deal with ``Reggeized" $t$-channel
partons.
If the particles produced in the collision are strongly separated in rapidity,
they obey multi-Regge kinematics (MRK).
If the same situation is realized with groups of particles, then
quasi-multi-Regge kinematics (QMRK) is at work.
In the case of single jet or prompt-photon inclusive production, this means the
following:
A single jet or a prompt photon is produced in the central region of rapidity,
while other particles are produced with large modula of rapidities.
In the experiment, the requirement of separation in rapidity can be controlled
by the so-called isolation cone condition.

Previously, in Ref.~\cite{Kim:1997dx}, single jet inclusive production was
studied in the Regge limit of QCD using the Balitsky-Fadin-Kuraev-Lipatov
(BFKL) framework \cite{BFKL}, and it was shown that the discrepancy between
data and theory in the region of small values of $x_T=2p_T/\sqrt S$ may be
accounted for by the BFKL Pomeron.
However, Pomeron exchange should be a dominant mechanism only at asymptotically
large energies.
In fact, in the energy range of the Tevatron and the LHC, the mechanism of
Reggeized gluon and quark exchange should be more adequate.

Later, in Ref.~\cite{Ostrovsky}, the infrared-stable single jet inclusive cross
section was calculated at next-to-leading order (NLO) in the framework of
high-energy factorization using the unintegrated gluon PDF of the asymptotic
BFKL approach and the one simply obtained by differentiating the collinear one
w.r.t.\ the scale parameter, in compliance with BFKL evolution.
The scatterings of off-shell partons were described by generalized cross
sections calculated in the QMRK approach \cite{QMRK}.
In contrast to the case of collinear factorization, NLO corrections were found
to diminish the single jet inclusive cross section in the framework of
high-energy factorization.
However, in Ref.~\cite{Ostrovsky}, the region of very small jet transverse
momentum, $p_T< 20$~GeV, was analyzed, which lies far below the $p_T$ range
studied experimentally at the Tevatron and the LHC, $p_T>50$~GeV, so that
the predictions of Ref.~\cite{Ostrovsky} cannot be tested.
Also taking into account that the unintegrated gluon PDFs are so far not well
constrained, we consider the result of Ref.~\cite{Ostrovsky} to be preliminary
and approximate.

The parton Reggeization framework \cite{QMRK} is particularly appropriate for
this kind of high-energy phenomenology.
It is based on an effective quantum field theory implemented with the
non-Abelian gauge-invariant action including fields of Reggeized
gluons \cite{Lipatov95} and Reggeized quarks \cite{LipatoVyazovsky}.

Recently, this approach was successfully applied to interpret the production of
prompt photons \cite{SVADISy},
diphotons \cite{SVAdiy},
charmed mesons \cite{PRD},
bottom-flavored jets \cite{PRb},
charmonia \cite{Kniehl:2006sk}, and
bottomonia \cite{Kniehl:2006vm}
as measured at the Tevatron and at DESY HERA in the small-$x_T$ regime.
In this paper, we continue our work in the parton Reggeization framework by
studying the distributions in transverse momentum and rapidity ($y$) of
single jet and prompt-photon inclusive hadroproduction.
We assume the MRK production mechanism to be the dominant one at small $x_T$
values.
We compare our results with experimental data taken by the CDF \cite{RRgCDF}
and D0 \cite{RRgD0,y18,y196} Collaborations at the Tevatron with
$\sqrt S=1.8$~TeV and 1.96~TeV and by the ATLAS Collaboration
\cite{Atlas,yAtlas} at the LHC with $\sqrt S=7$~TeV.
We also present predictions for the $p_T$ and $y$ distributions of single jet
and prompt-photon inclusive production at the LHC with $\sqrt{S}=14$~TeV.

\section{Born amplitudes with multi-Regge kinematics}

We examine single jet and prompt-photon inclusive production in
proton-antiproton collisions at the Tevatron and in
proton-proton collisions at the LHC.
To leading order (LO) in the parton Reggeization framework, the relevant
hard-scattering processes are $\mathcal{R}+\mathcal{R}\to g$ and
$\mathcal{Q}+\overline{\mathcal{Q}}\to \gamma$, where $\mathcal{R}$ is a
Reggeized gluon, $g$ is a Yang-Mills gluon, $\mathcal{Q}$ is a Reggeized
quark, and $\gamma$ is a photon.
Working in the center-of-mass (c.m.) frame, we write the four-momenta of the
incoming hadrons as $P_{1,2}^\mu=(\sqrt{S}/2)(1,0,0,\pm1)$ and those of the
Reggeized partons as $q_i^\mu=x_iP_i^\mu+q_{iT}^\mu$ ($i=1,2)$, where $x_i$ are the
longitudinal momentum fractions and $q_{iT}^\mu=(0,\mathbf{q}_{iT},0)$, with
$\mathbf{q}_{iT}$ being transverse two-momenta, and we define
$t_i=-q_{iT}^2=\mathbf{q}_{iT}^2$.
The gluon and photon produced in the $2\to1$ partonic subprocesses have
four-momentum $p^\mu=q_1^\mu+q_2^\mu=(p^0,\mathbf{p}_T,p^3)$, with
$\mathbf{p}_T^2=t_1+t_2+2\sqrt{t_1t_2}\cos\phi_{12}$, where
$\phi_{12}$ is the azimuthal angle enclosed between
$\mathbf{q}_{1T}$ and $\mathbf{q}_{1T}$.
Introducing the light-cone vectors $(n^\pm)^\mu=(1,0,0,\pm1)$, we define
$k^{\pm}=k\cdot n^\pm$ for any four-vector $k^\mu$.

The Fadin-Kuraev-Lipatov effective $\mathcal{R}\mathcal{R}g$
vertex reads \cite{BFKL,RRgold}:
\begin{equation}
C_{\mathcal{RR}}^{g,\mu}(q_1,q_2)=-\sqrt{4\pi\alpha_s}f^{abc}
\frac{q_1^+q_2^-}{2\sqrt{t_1t_2}}
\left[\left(q_1-q_2\right)^\mu+
\frac{(n^+)^\mu}{q_1^+}\left(q_2^2+q_1^+q_2^-
\right)-
\frac{(n^-)^\mu}{q_2^-}\left(q_1^2+q_1^+q_2^-\right)\right],
\label{amp:RRg}
\end{equation}
where $\alpha_s$ is the strong-coupling constant, $a$ and $b$ are the color
indices of the Reggeized gluons with incoming four-momenta $q_1$ and $q_2$, and
$f^{abc}$ are the structure constants of the color group group SU(3).
The squared amplitude of the partonic subprocess
$\mathcal{R}+\mathcal{R}\to g$ is straightforwardly found from
Eq.~(\ref{amp:RRg}) to be
\begin{equation}
\overline{|{\cal M}(\mathcal{R}+\mathcal{R}\to
g)|^2}=\frac{3}{2}\pi \alpha_s \mathbf{p}_T^2.
\label{sqamp:RRg}
\end{equation}

Neglecting quark masses, the effective
$\mathcal{Q}\overline{\mathcal{Q}}\gamma$ and
$\mathcal{Q}\overline{\mathcal{Q}}g$ vertices read \cite{QR1old}:
\begin{equation}
C_{\mathcal{Q}\overline{\mathcal{Q}}}^{\gamma/g,\mu}(q_1,q_2)=C_1^{\gamma/g}
\left[\gamma^\mu-\slashed{q}_1\frac{(n^-)^\mu}{q^-_1+q^-_2}-\slashed{q}_2
\frac{(n^+)^\mu}{q^+_1+q^+_2}\right],
\label{amp:QQg}
\end{equation}
where $C_1^\gamma=-i\sqrt{4\pi\alpha}e_q$, with $\alpha$ being Sommerfeld's
fine-structure constant and $e_q$ being the fractional charge of quark $q$ (and
its Reggeized variant $\mathcal{Q}$), and $C_1^g=-i\sqrt{4\pi\alpha_s}T^a$,
with $T^a$ being a generator of SU(3).
The squared amplitudes of the partonic subprocesses
$\mathcal{Q}\overline{\mathcal{Q}}\to\gamma$ and
$\mathcal{Q}\overline{\mathcal{Q}}\to g$ are found from Eq.~(\ref{amp:QQg}) to
be
\begin{equation}
\overline{\left|{\cal M}\left(\mathcal{Q}+\overline{\mathcal{Q}}\to
\gamma/g\right)\right|^2}=C_2^{\gamma/g}(t_1+t_2),
\label{sqamp:QQg}
\end{equation}
where $C_2^\gamma=(4/3)\pi\alpha e_q^2$ and
$C_2^g=(16/3)\pi\alpha_s$.

\section{Cross sections}

Exploiting the hypothesis of high-energy factorization, we may write the
hadronic cross sections $d\sigma$ as convolutions of partonic cross sections
$d\hat\sigma$ with unintegrated PDFs $\Phi_a^h$ of Reggeized partons $a$ in the
hadrons $h$, as
\begin{equation}
d\sigma\left(p\overline{p}\to j+X\right)=\int\frac{dx_1}{x_1}\int
\frac{d^2q_{1T}}{\pi}\int\frac{dx_2}{x_2}\int\frac{d^2q_{2T}}{\pi}
\Phi^p_{g}(x_1,t_1,\mu^2)\Phi^{\overline{p}}_{g}(x_2,t_2,\mu^2)
d\hat\sigma\left(\mathcal{R}\mathcal{R}\to g\right),
\label{eq:XSRRg}
\end{equation}
and similarly for $pp$ collisions and single prompt-photon production.
For the reader's convenience, we also present here a compact formula for the
double differential distribution in $p_T=|\mathbf{p_T}|$ and $y$, which follows
from Eq.~(\ref{eq:XSRRg}) and reads:
\begin{equation}
\frac{d\sigma}{dp_T\,dy}\left(p\overline{p}\to j+X\right)
=\frac{1}{p_T^3}\int d\phi_1\int dt_1
\Phi^p_g(x_1,t_1,\mu^2)\Phi^{\overline{p}}_g(x_2,t_2,\mu^2)
\overline{\left|{\cal M}\left(\mathcal{R}\mathcal{R}\to g\right)\right|^2},
\label{eq:cXSRRg}
\end{equation}
where $\phi_1$ is the azimuthal angle enclosed between $\mathbf{q}_{1T}$ and
$\mathbf{p}_T$,
\begin{equation}
x_{1,2}=\frac{p_T\exp(\pm y)}{\sqrt{S}},\qquad
t_2=t_1+p_T^2-2p_T\sqrt{t_1}\cos\phi_1.
\end{equation}
In the case of single prompt-photon inclusive production, we take the first
three quark flavors, $u$, $d$, and $s$, to be active.
Since we work at LO, the produced jet has zero invariant mass $m$, so that
transverse energy $E_T=\sqrt{p_T^2+m^2}$ and transverse momentum $p_T$ coincide
and so do rapidity $y=(1/2)\ln[(p^0+p^3)/(p^0-p^3)]$ and pseudorapidity
$\eta=-\ln\tan(\theta/2)$, where $\theta$ is the angle enclosed between the
jet and beam axes.

The unintegrated PDFs $\Phi_a^h(x,t,\mu^2)$ are related to their
collinear counterparts $F_a^h(x,\mu^2)$ by the normalization
condition,
\begin{equation}
xF_a^h(x,\mu^2)=\int^{\mu^2}dt\,\Phi_a^h(x,t,\mu^2),
\end{equation}
which yields the correct transition from formulas in the QMRK approach to
those in the collinear parton model, where the transverse momenta
of the partons are neglected.
In our numerical analysis, we adopt as our default the prescription proposed by
Kimber, Martin, and Ryskin (KMR) \cite{KMR} to obtain unintegrated gluon and
quark PDFs of the proton from the conventional integrated ones, as implemented
in Watt's code \cite{Watt}.
As is well known \cite{Andersson:2002cf}, other popular prescriptions, such as
those by Bl\"umlein (B) \cite{Blumlein:1995eu} or by Jung and Salam
\cite{Jung:2000hk}, produce unintegrated PDFs with distinctly different $t$
dependences.
In order to assess the resulting theoretical uncertainty, we also evaluate the
unintegrated gluon PDF using the B approach, which resums small-$x$ effects
according to the BFKL equation. 
As input for these procedures, we use the LO set of the
Martin-Roberts-Stirling-Thorne (MRST) \cite{MRST} proton PDFs as our default.
In order to estimate the theoretical uncertainty due to the freedom in the
choice of the PDFs, we also use the CTEQ6L1 set by the CTEQ Collaboration
\cite{Pumplin:2002vw} as well as the Gl\"uck-Reya-Vogt (GRV)
\cite{Gluck:1994uf} LO set.

Throughout our analysis the renormalization and factorization scales are
identified and chosen to be $\mu=\xi p_T$, where $\xi$ is varied between 1/2
and 2 about its default value 1 to estimate the theoretical uncertainty due to
the freedom in the choice of scales.
The resulting errors are indicated as shaded bands in the figures.

\section{Results}

\begin{figure}[ht]
\begin{center}
\includegraphics[width=.8\textwidth, clip=]{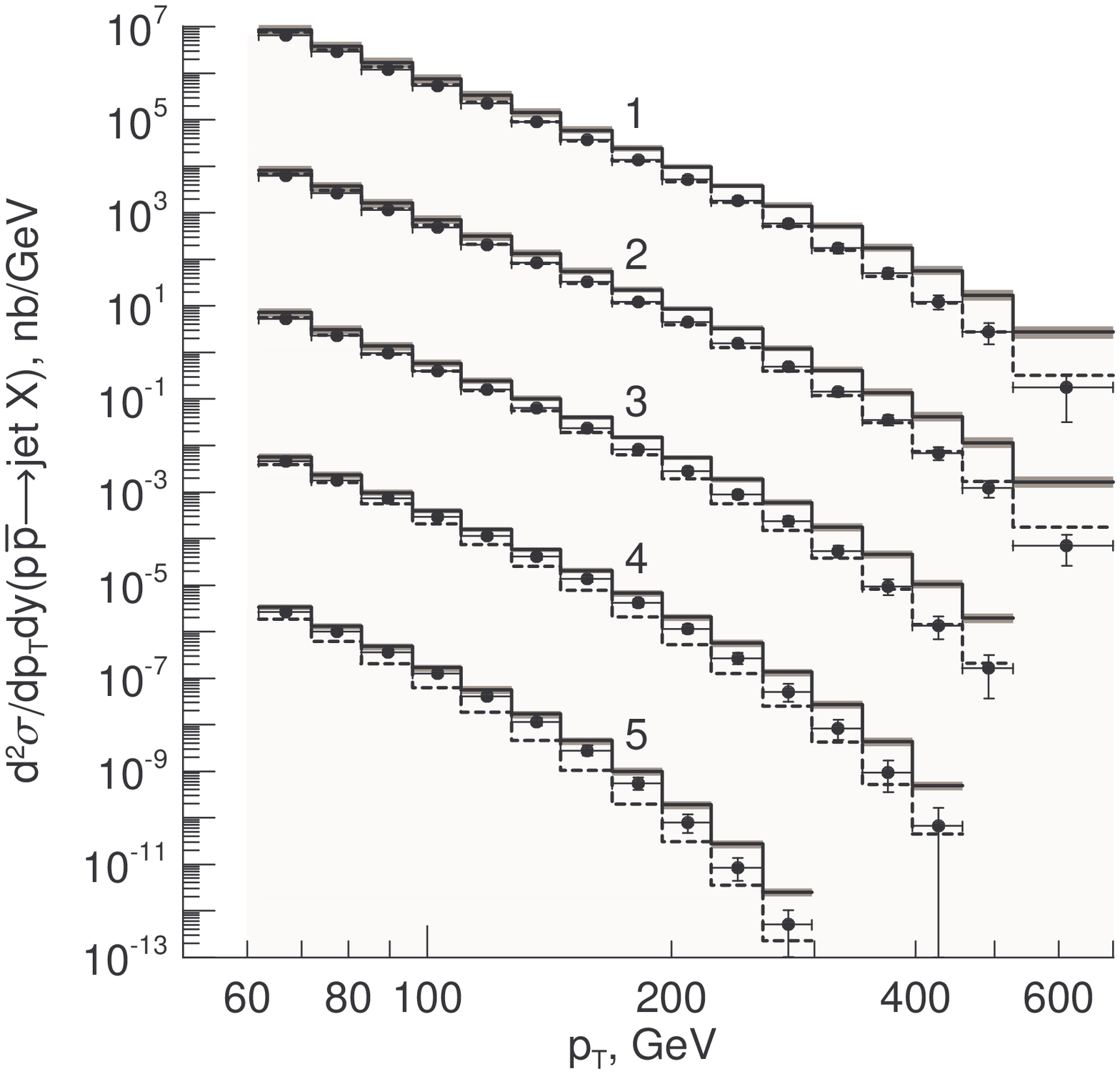}
\end{center}
\caption{\label{fig:1}%
The transverse-momentum distributions of single jet inclusive
hadroproduction measured in the rapidity intervals
(1) $|y|<0.1$ (${}\times 10^{6}$),
(2) $0.1<|y|<0.7$ (${}\times 10^{3}$),
(3) $0.7<|y|<1.1$,
(4) $1.1<|y|<1.6$ (${}\times 10^{-3}$), and
(5) $1.6<|y|<2.1$ (${}\times 10^{-6}$)
by the CDF Collaboration in Tevatron run~II \cite{RRgCDF} are compared with
our LO MRK predictions evaluated in the KMR (solid histograms) and B (dashed
histograms) approaches using the MRST PDFs. 
The shaded bands indicate the scale uncertainties in the KMR evaluations.}
\end{figure}

\begin{figure}[ht]
\begin{center}
\includegraphics[width=.8\textwidth, clip=]{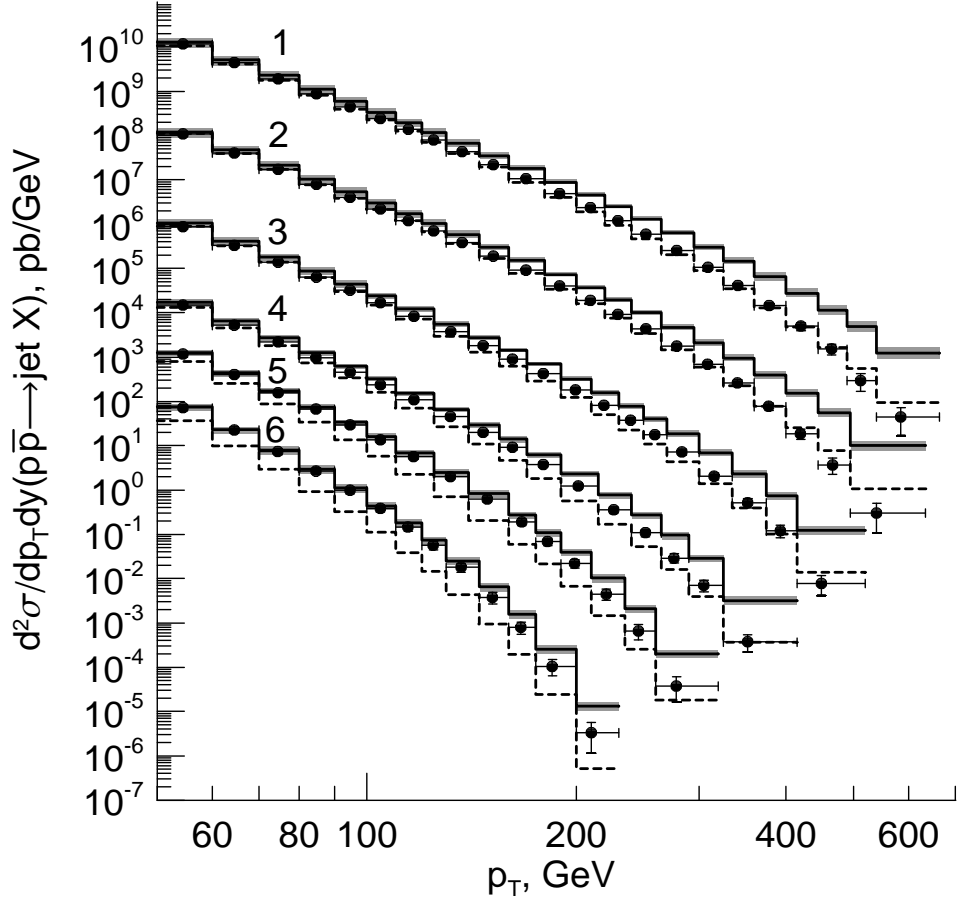}
\end{center}
\caption{\label{fig:2}%
The transverse-momentum distributions of single jet inclusive
hadroproduction measured in the rapidity intervals
(1) $|y|<0.4$ (${}\times 5\cdot10^{5}$),
(2) $0.4<|y|<0.8$ (${}\times 5\cdot 10^{3}$),
(3) $0.8<|y|<1.2$ (${}\times 50$),
(4) $1.2<|y|<1.6$,
(5) $1.6<|y|<2.0$ (${}\times 0.1$), and
(6) $2.0<|y|<2.4$ (${}\times 10^{-2}$)
by the D0 Collaboration in Tevatron run~II \cite{RRgD0} are compared with
our LO MRK predictions evaluated in the KMR (solid histograms) and B (dashed
histograms) approaches using the MRST PDFs. 
The shaded bands indicate the scale uncertainties in the KMR approach.}
\end{figure}

We are now in a position to present our theoretical predictions and to
compare them with experimental measurements.
We first consider single jet inclusive production.
Recently, the CDF \cite{RRgCDF} (D0 \cite{RRgD0}) Collaboration presented new
data from Tevatron run~II, which correspond to an integrated luminosity of
1.13~fb$^{-1}$ (0.70~fb$^{-1}$) and cover the kinematic range
62~GeV${}<p_T<700$~GeV (50~GeV${}<p_T<600$~GeV) and $|y|<2.1$ ($|y|<2.4$).
The CDF and D0 data are compared with our MRK predictions in Figs.~\ref{fig:1}
and \ref{fig:2}, respectively.
We find good agreement for $p_T\alt100$~GeV, which corresponds to
$x_T\alt0.1$, while our default predictions overshoot the data for larger
values of $p_T$.
This may be understood by observing that the average values of the scaling
variables $x_1$ and $x_2$ in Eq.~(\ref{eq:cXSRRg}) are of order $x_T$, and
the MRK picture ceases to be valid for $x_i\agt0.1$.
For $x_T\agt0.1$, one needs to resort to the collinear parton model, which
starts with $2\to2$ partonic subprocesses at LO.
Since the unintegrated quark PDFs are greatly suppressed compared to the gluon
one, the contributions due to partonic subprocesses involving Reggeized quarks,
such as $\mathcal{R}\mathcal{Q}\to q$ and
$\mathcal{Q}\overline{\mathcal{Q}}\to g$, are expected to be relatively small
in the relevant $x_T$ range, $x_T\agt0.1$.
The predictions obtained using the B approach undershoot the default ones
leading to a better description of the experimental data at large values of
$p_T$, where the MRK picture is, however, not expected to apply.
However, they undershoot the experimental data at large values of $|y|$
throughout the whole $p_T$ range considered.

\begin{figure}[ht]
\begin{center}
\includegraphics[width=.8\textwidth, clip=]{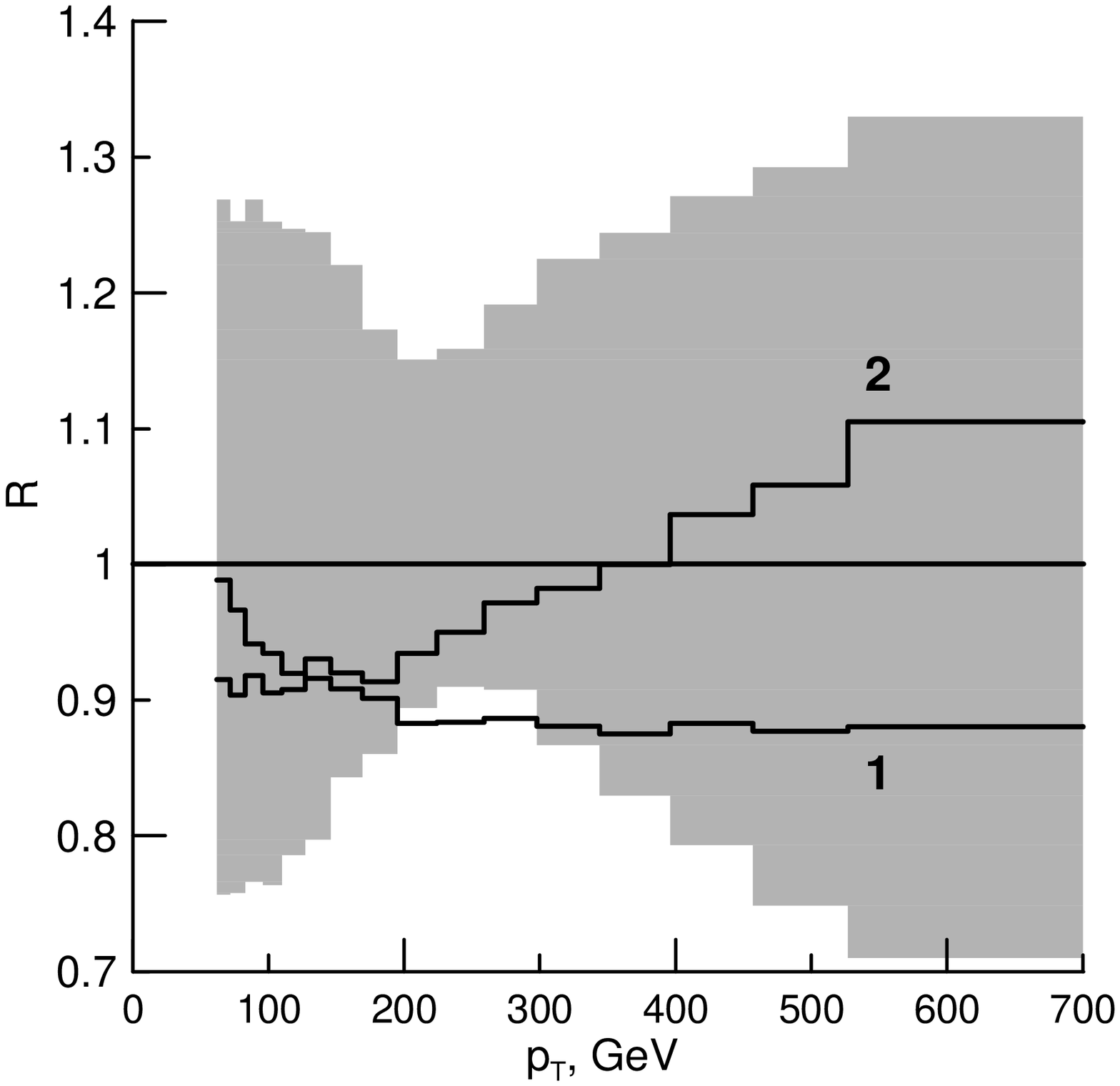}
\end{center}
\caption{\label{fig:uipdf1}%
Theoretical uncertainties in the KMR approach due to the freedom in the choices
of scales and unintegrated PDF set in the transverse-momentum distribution of
single jet inclusive hadroproduction in $p\overline{p}$ collisions with
$\sqrt{S}=1.96$~TeV integrated over the rapidity interval $|y|<0.1$.
The evaluation with the MRST LO set and $\xi$ varied in the interval
$1/2<\xi<2$ (shaded band) and those with the (1) CTEQ6L1 and (2) GRV LO sets
and $\xi=1$ are normalized to the one with the MRST LO set and $\xi=1$.}
\end{figure}

\begin{figure}[ht]
\begin{center}
\includegraphics[width=.8\textwidth, clip=]{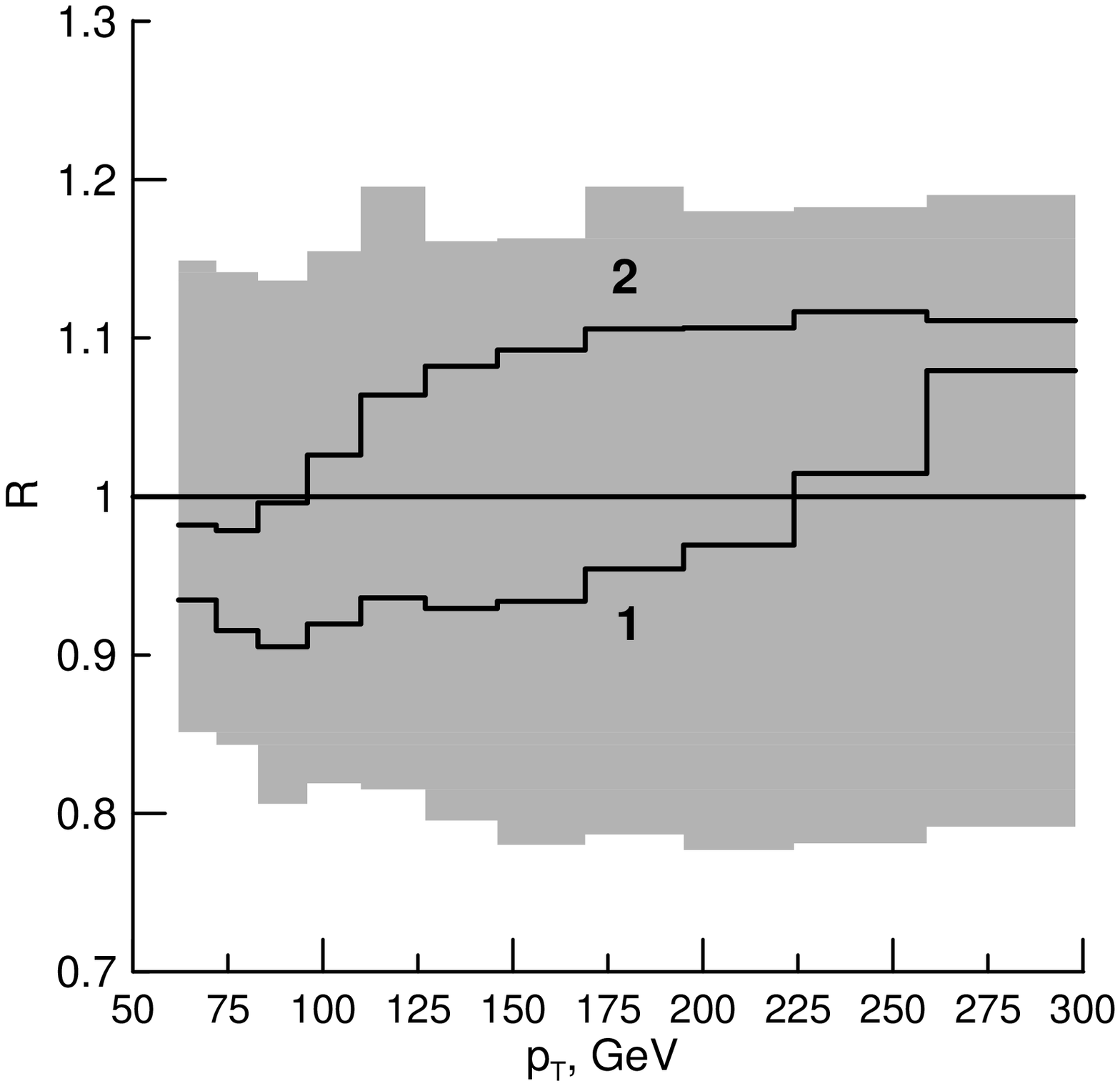}
\end{center}
\caption{\label{fig:uipdf2}%
Theoretical uncertainties in the KMR approach due to the freedom in the choices
of scales and unintegrated PDF set in the transverse-momentum distribution of
single jet inclusive hadroproduction in $p\overline{p}$ collisions with
$\sqrt{S}=1.96$~TeV integrated over the rapidity interval $1.6<|y|<2.1$.
The evaluation with the MRST LO set and $\xi$ varied in the interval
$1/2<\xi<2$ (shaded band) and those with the (1) CTEQ6L1 and (2) GRV LO sets
and $\xi=1$ are normalized to the one with the MRST LO set and $\xi=1$.}
\end{figure}

In Figs.~\ref{fig:1} and \ref{fig:2}, the theoretical uncertainties due to the
fredom in the choices of the renormalization and factorization scales are
indicated for the default predictions by the shaded bands.
Our limited knowledge of the unintegrated PDFs also contributes to the
theoretical uncertainty.
In Figs.~\ref{fig:uipdf1} and \ref{fig:uipdf2}, we investigate this source
of theoretical uncertainty for the $p_T$ distribution of single jet
inclusive hadroproduction in $p\overline{p}$ collisions with
$\sqrt{S}=1.96$~TeV integrated over the rapidity intervals $|y|<0.1$ and
$1.6<|y|<2.1$, respectively.
Specifically, we consider the evaluations, for $\xi=1$, with the CTEQ6L1
\cite{Pumplin:2002vw} and GRV LO \cite{Gluck:1994uf} PDFs normalized to the
ones with the MRST LO PDFs \cite{MRST}.
These ratios are typically well contained within the bands generated by
varying $\xi$ between 1/2 and 2 in the evaluations with the MRST LO PDFs.
We thus conclude that scale variations provide reasonable estimates of the
overall theoretical uncertainties in the evaluations based on the KMR approach.

\begin{figure}[ht]
\begin{center}
\includegraphics[width=.8\textwidth, clip=]{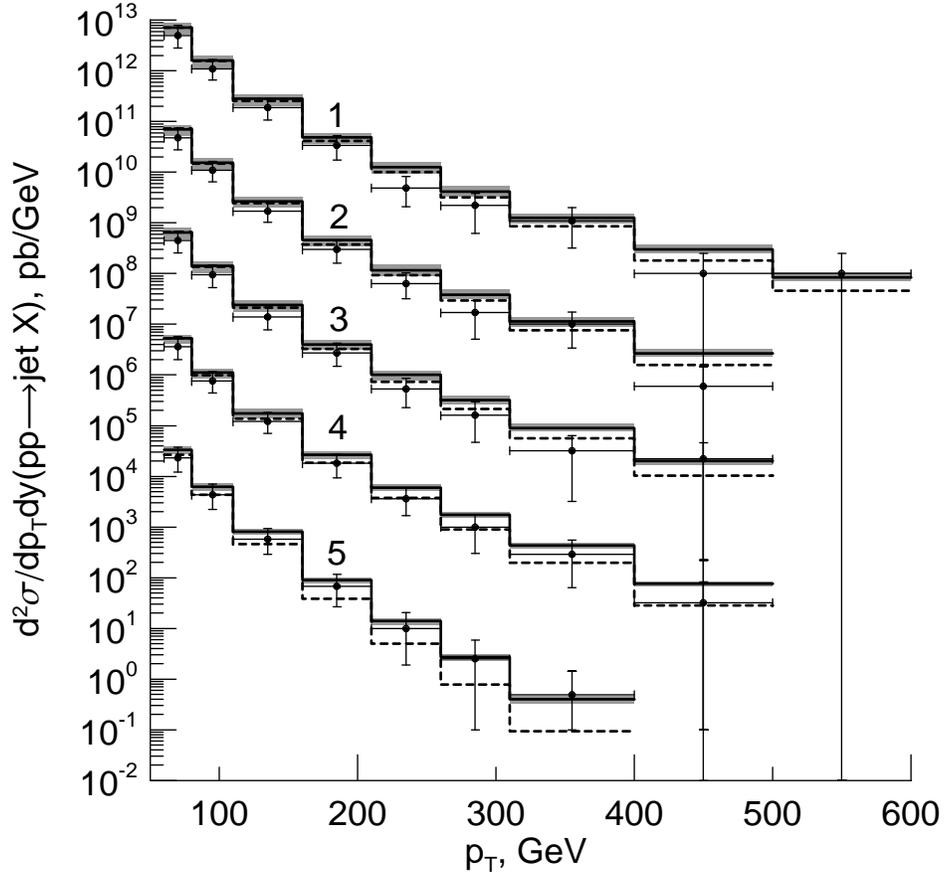}
\end{center}
\caption{\label{fig:3}%
The transverse-momentum distributions of single jet inclusive
hadroproduction measured in the rapidity intervals
(1) $|y|<0.3$ (${}\times 10^{8}$),
(2) $0.3<|y|<0.8$ (${}\times 10^{6}$),
(3) $0.8<|y|<1.2$ (${}\times 10^{4}$),
(4) $1.2<|y|<2.1$ (${}\times 10^{2}$), and
(5) $2.1<|y|<2.6$
by the ATLAS Collaboration at the LHC \cite{Atlas} are compared with
our LO MRK predictions evaluated in the KMR (solid histograms) and B (dashed
histograms) approaches using the MRST PDFs. 
The shaded bands indicate the scale uncertainties in the KMR evaluations.}
\end{figure}

\begin{figure}[ht]
\begin{center}
\includegraphics[width=.8\textwidth, clip=]{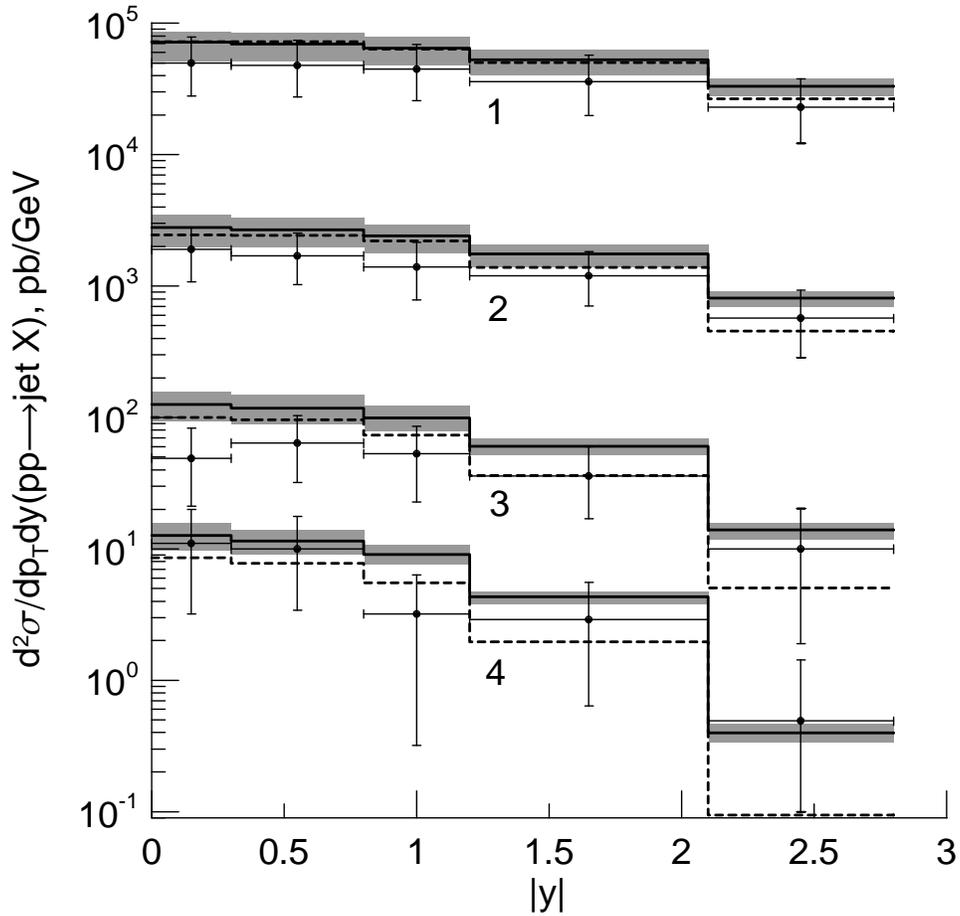}
\end{center}
\caption{\label{fig:4}%
The rapidity distributions of single jet inclusive hadroproduction measured in
the transverse-momentum intervals
(1) 60~GeV${}<p_T<80$~GeV,
(2) 110~GeV${}<p_T<160$~GeV,
(3) 210~GeV${}<p_T<250$~GeV, and
(4) 310~GeV${}<p_T<400$~GeV
by the ATLAS Collaboration at the LHC \cite{Atlas} are compared with
our LO MRK predictions evaluated in the KMR (solid histograms) and B (dashed
histograms) approaches using the MRST PDFs. 
The shaded bands indicate the scale uncertainties in the KMR evaluations.}
\end{figure}

Moving on from the Tevatron to the LHC, which is currently running at
$\sqrt S=7$~TeV, being about 3.5 larger than at the Tevatron, one expects
the $p_T$ range of validity of the MRK picture to be extended by the same
factor, to $p_T\alt350$~GeV.
This expectation is nicely confirmed in Figs.~\ref{fig:3} and \ref{fig:4},
where a recent measurement by the ATLAS Collaboration \cite{Atlas}, which is
based on an integrated luminosity of 17~nb$^{-1}$ and covers the kinematic
range 60~GeV${}<p_T<600$~GeV and $|y|<2.8$, is compared with our MRK
predictions for the $p_T$ and $y$ distributions, respectively.
In fact, useful agreement is found even through the largest $p_T$ values
accessed by this measurement.

Note that, in Ref.~\cite{Atlas}, jets are identified using the anti-$k_t$
jet-clustering algorithm with two different values of the jet-size parameter
$R=\sqrt{(\Delta y)^2+(\Delta\phi)^2}$, namely $R=0.4$ and $R=0.6$.
The ATLAS data shown in Figs.~\ref{fig:3} and \ref{fig:4} refer to $R=0.6$.
The agreement is somewhat worse for $R=0.4$.
This may be understood by observing that the MRK picture assumes a strong
hierarchy in $y$ and thus prefers a stronger isolation.
Our LO prediction does not yet depend on $R$.

\begin{figure}[ht]
\begin{center}
\includegraphics[width=.8\textwidth, clip=]{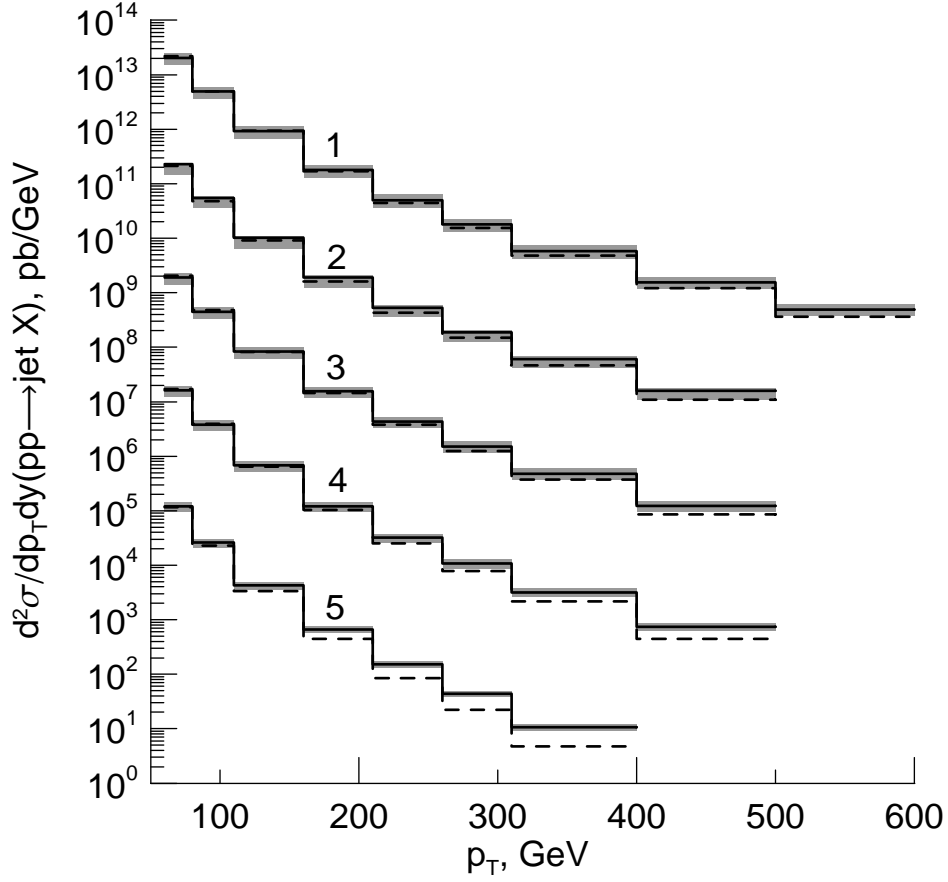}
\end{center}
\caption{\label{fig:5}%
Transverse-momentum distributions of single jet inclusive hadroproduction in
$pp$ collisions with $\sqrt{S}=14$~TeV integrated over the rapidity intervals
(1) $|y|<0.3$ ($\times 10^{8}$),
(2) $0.3<|y|<0.8$ (${}\times 10^{6}$),
(3) $0.8<|y|<1.2$ (${}\times 10^{4}$),
(4) $1.2<|y|<2.1$ (${}\times 10^{2}$), and
(5) $2.1<|y|<2.6$
as predicted at LO in the MKR framework adopting the KMR (solid histograms) and
B (dashed histograms) approaches with the MRST PDFs. 
The shaded bands indicate the scale uncertainties in the KMR evaluations.}
\end{figure}

\begin{figure}[ht]
\begin{center}
\includegraphics[width=.8\textwidth, clip=]{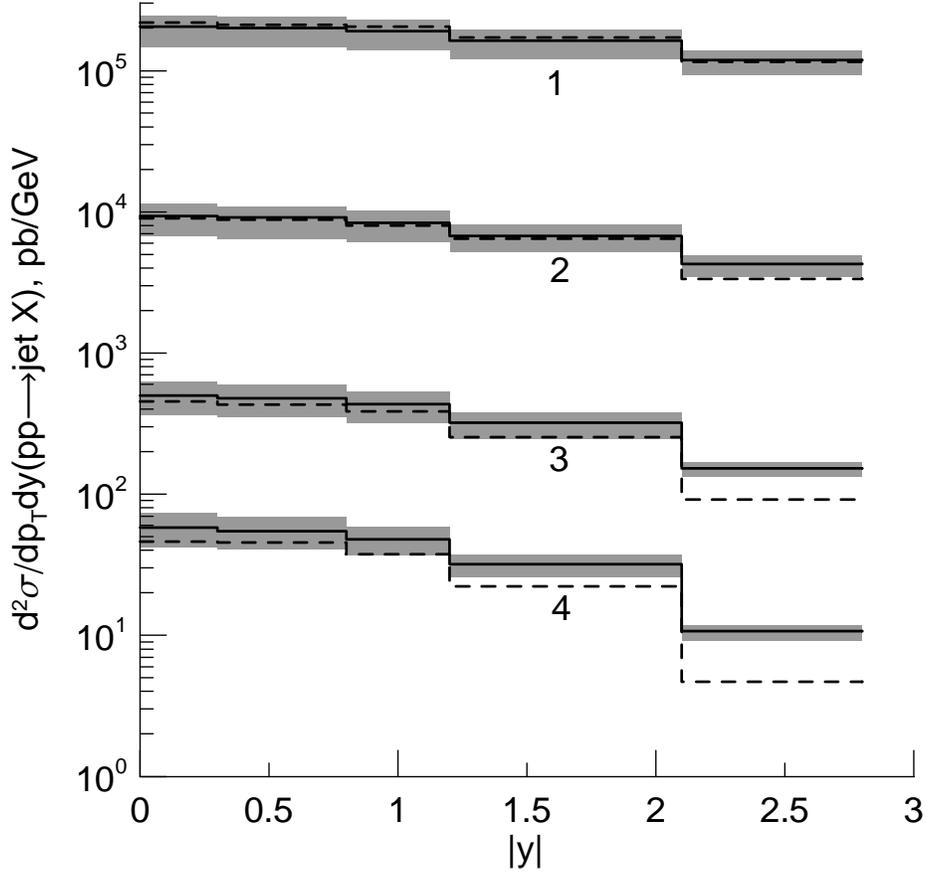}
\end{center}
\caption{\label{fig:6}%
Rapidity distributions of single jet inclusive hadroproduction in
$pp$ collisions with $\sqrt{S}=14$~TeV integrated over the
transverse-momentum intervals
(1) 60~GeV${}<p_T<80$~GeV,
(2) 110~GeV${}<p_T<160$~GeV,
(3) 210~GeV${}<p_T<250$~GeV, and
(4) 310~GeV${}<p_T<400$~GeV
as predicted at LO in the MKR framework adopting the KMR (solid histograms) and
B (dashed histograms) approaches with the MRST PDFs. 
The shaded bands indicate the scale uncertainties in the KMR evaluations.}
\end{figure}

In Figs.~\ref{fig:5} and \ref{fig:6}, we repeat the MRK analyses of
Figs.~\ref{fig:3} and \ref{fig:4} for the LHC design c.m.\ energy
$\sqrt{S}=14$~TeV, where we expect the $p_T$ range of validity to be roughly
$p_T\alt700$~GeV.

\begin{figure}[ht]
\begin{center}
\includegraphics[width=.8\textwidth, clip=]{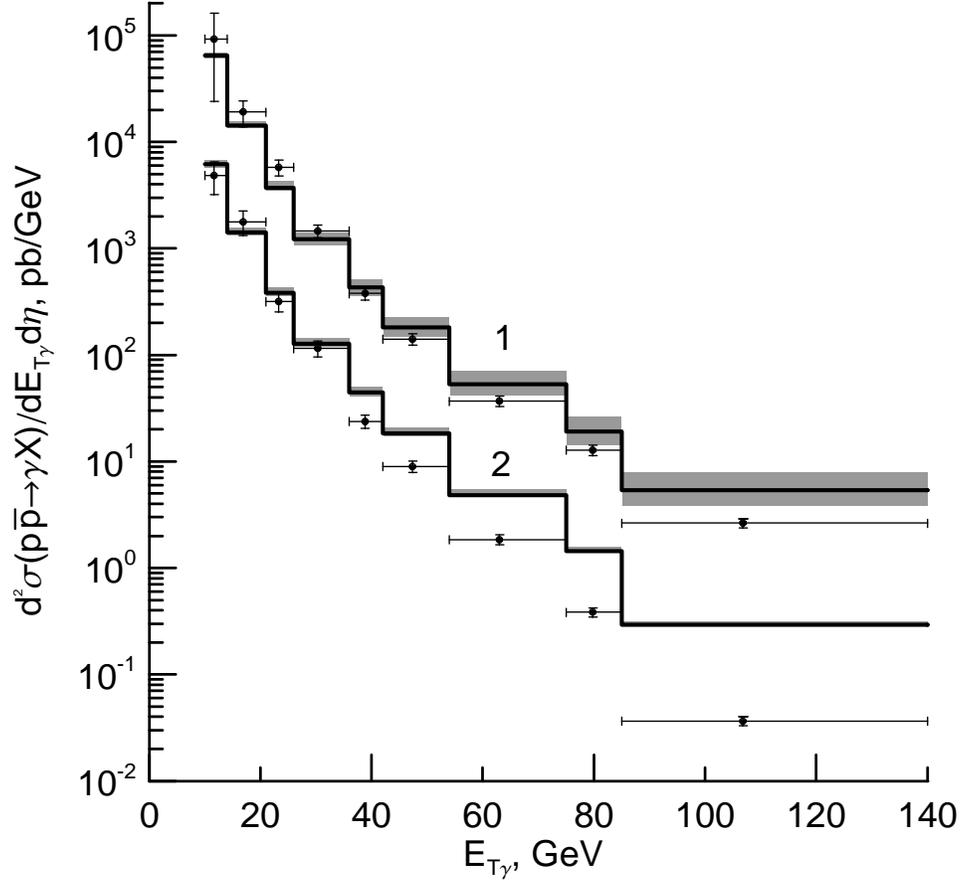}
\end{center}
\caption{\label{fig:7}%
The transverse-energy distributions of single prompt-photon inclusive
hadroproduction measured in the pseudorapidity intervals
(1) $|\eta|<0.9$ (${}\times 10$) and
(2) $1.6<|\eta|<2.5$
by the D0 Collaboration in Tevatron run~I \cite{y18} are compared with
our LO MRK predictions.
The shaded bands indicate the theoretical uncertainties.}
\end{figure}

\begin{figure}[ht]
\begin{center}
\includegraphics[width=.8\textwidth, clip=]{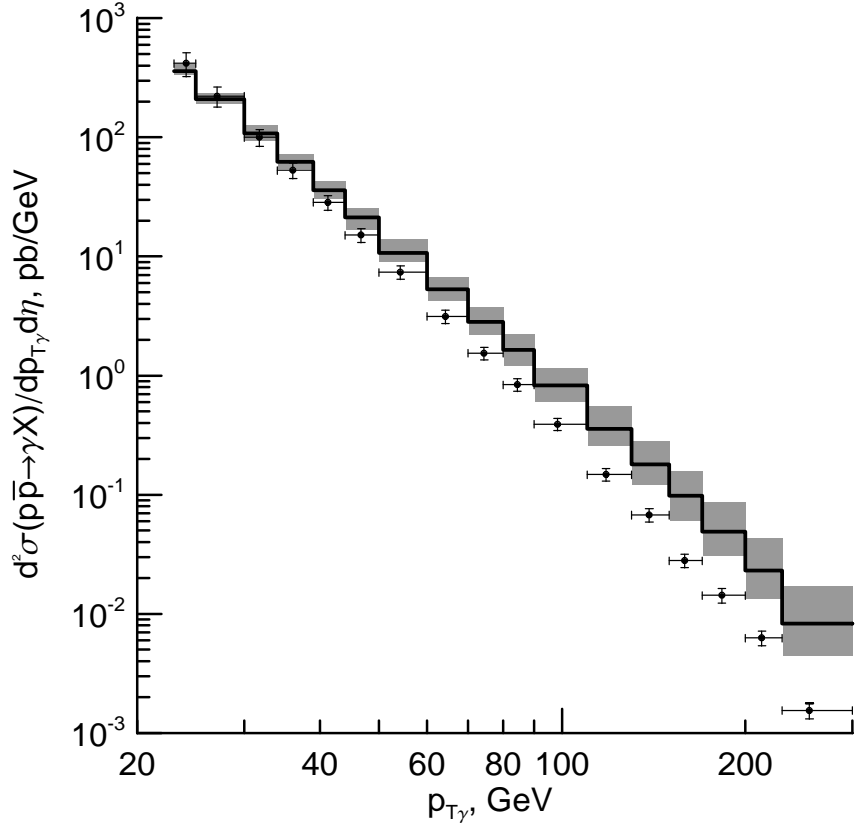}
\end{center}
\caption{\label{fig:8}%
The transverse-momentum distribution of single prompt-photon inclusive
hadroproduction measured in the pseudorapidity interval $|\eta|<0.9$
by the D0 Collaboration in Tevatron run~II \cite{y196} is compared with
our LO MRK predictions.
The shaded band indicates the theoretical uncertainty.}
\end{figure}

Let us now turn to single prompt-photon inclusive production.
In Figs.~\ref{fig:7} and \ref{fig:8}, we compare our MRK predictions with
data taken by the D0 Collaboration in Tevatron runs~I \cite{y18} and II
\cite{y196}, respectively.
The analysis of Ref.~\cite{y18} (\cite{y196}) is based on an integrated
luminosity of 107.6~pb$^{-1}$ (326~pb$^{-1}$) and covers the kinematic range
10~GeV${}<E_T<140$~GeV (23~GeV${}<p_T<300$~GeV) and $|\eta|<2.5$
($|\eta|<0.9$).
We find reasonable agreement through $E_T\approx85$~GeV ($p_T\approx60$~GeV)
for the central events, with $|\eta|<0.9$, from run~I \cite{y18} (run~II
\cite{y196}), but only through $p_T\approx36$~GeV for the forward events, with
$1.6<|\eta|<0.9$, from run~I \cite{y18}.
Fragmentation production, via partonic subprocesses such as
$\mathcal{R}\mathcal{Q}\to q\to\gamma$ and
$\mathcal{R}\mathcal{R}\to g\to\gamma$, should be numerically small compared to
direct production and is neglected in our exploratory analysis.

\begin{figure}[ht]
\begin{center}
\includegraphics[width=.8\textwidth, clip=]{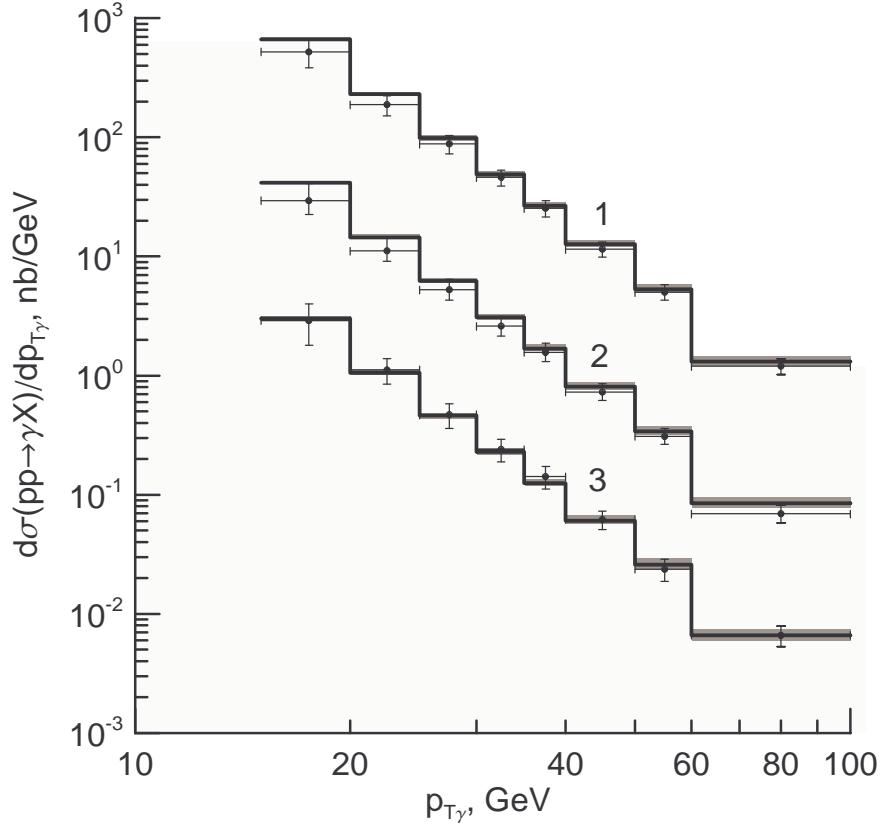}
\end{center}
\caption{\label{fig:9}%
The transverse-momentum distributions of single prompt-photon inclusive
hadroproduction measured in the pseudorapidity intervals
(1) $|\eta|<0.6$ (${}\times 10^2$),
(2) $0.6<|\eta|<1.37$ (${}\times 5$), and
(3) $1.52<|\eta|<1.81$
by the ATLAS Collaboration at the LHC \cite{Atlas} are compared with
our LO MRK predictions.
The shaded bands indicate the theoretical uncertainties.}
\end{figure}

In Fig.~\ref{fig:9}, we compare our MRK predictions with a very recent
measurement by the ATLAS Collaboration, which is based on an integrated
luminosity of 880~nb$^{-1}$ and covers the kinematic range
15~GeV${}<p_T<100$~GeV and $|\eta|<1.81$.
The agreement is found to be excellent, as expected because of the small $x_T$
values probed.

\begin{figure}[ht]
\begin{center}
\includegraphics[width=.8\textwidth, clip=]{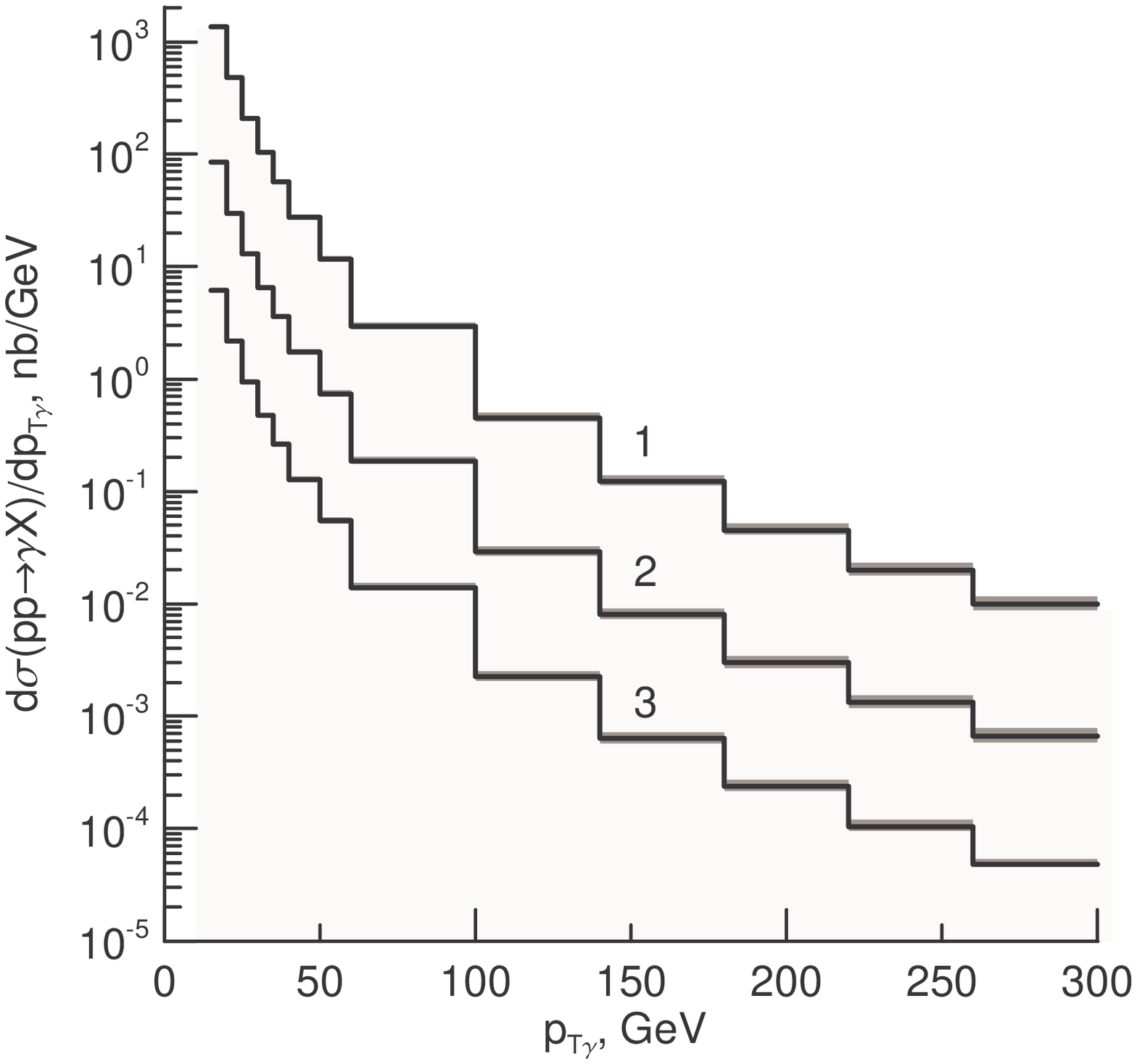}
\end{center}
\caption{\label{fig:10}%
Transverse-momentum distributions of single prompt-photon inclusive
hadroproduction in $pp$ collisions with $\sqrt{S}=14$~TeV integrated over the
pseudorapidity intervals
(1) $|\eta|<0.6$ (${}\times 10^2$),
(2) $0.6<|\eta|<1.37$ (${}\times 5$), and
(3) $1.52<|\eta|<1.81$
as predicted at LO in the MKR approach.
The shaded bands indicate the theoretical uncertainties.}
\end{figure}

Finally, we repeat the MRK analyses of Fig.~\ref{fig:9} for $\sqrt{S}=14$~TeV
and show the results in Fig.~\ref{fig:10}.

\section{Conclusions}
\label{sec:five}

The Tevatron and, even more so, the LHC are currently probing particle physics
at terascale c.m.\ energies $\sqrt{S}$, so that the hierarchy
$\Lambda_\mathrm{QCD}\ll\mu\ll\sqrt{S}$, which defines the MRK regime, is
satisfied for a wealth of QCD processes of typical energy scale $\mu$.

In this paper, we studied two QCD processes of particular interest, namely
single jet and prompt-photon inclusive hadroproduction, at LO in the MRK
approach, in which they are mediated by $2\to1$ partonic subprocesses
initiated by Reggeized gluons and quarks, respectively.
Despite the great simplicity of our analytic expressions, we found excellent
agreement with single jet \cite{Atlas} and prompt-photon \cite{yAtlas} data
taken just recently by the ATLAS Collaboration in $pp$ collisions with
$\sqrt{S}=7$~TeV at the LHC.
By contrast, in the collinear parton model of QCD, it is necessary to take
into account NLO corrections and to perform soft-gluon resummation in order
to obtain a comparable degree of agreement with the data, both for jet
\cite{Aversa:1988fv} and prompt-photon \cite{Aurenche:1987fs} inclusive
production.
However, our findings have to be taken with a grain of salt, since our LO
approach does not yet accommodate the concepts of single-jet cone radius and
prompt-photon isolation cone and neglects fragmentation to prompt photons.

On the other hand, comparisons with data taken by the CDF and D0
Collaborations at the Tevatron in $p\overline{p}$ collisons with
$\sqrt{S}=1.8$~TeV and 1.96~TeV, which is roughly a factor of 3.5 below the
value presently reached by the LHC, disclosed the limits of applicability of
the MRK picture.
In fact, the MRK approximation appears to break down for $x_T\agt0.1$ in the
case of single jet production and somewhat below that in the case of single
prompt-photon production.

These findings are in line with our previous studies of the MRK approach,
applied to the production of
prompt photons \cite{SVADISy},
diphotons \cite{SVAdiy},
charmed mesons \cite{PRD},
bottom-flavored jets \cite{PRb},
charmonia \cite{Kniehl:2006sk}, and
bottomonia \cite{Kniehl:2006vm}.
Here and in Refs.~\cite{SVADISy,SVAdiy,PRD,PRb,Kniehl:2006sk,Kniehl:2006vm},
parton Reggeization was demonstrated to be a powerful tool for the theoretical
description of QCD processes in the high-energy limit.

\section{Acknowledgements}

We are grateful to V.~S~Fadin and L.~N.~Lipatov for useful
discussions. This research was supported in part by the German
Federal Ministry for Education and Research BMBF under Grant No.\
05~HT6GUA, by the Helmholtz Association HGF under Grant No.\ Ha~101,
by the National Science Foundation NSF under Grant No.\
NSF~PHY05-51164, and by the Claussen-Simon-Stiftung. The work of
V.A.S. and A.V.S. was supported in part by the Federal Ministry for
Science and Education of the Russian Federation under Contract
No.~14.740.11.0894. A.V.S. is grateful to the International Center
of Fundamental Physics in Moscow and the Dynastiya Foundation for
financial support. The work of E.V.Y. was supported in part by
Michail Lomonosov Grant No.~A/09/72753, jointly funded by the German
Academic Exchange Service DAAD and the Ministry of Science and
Education of the Russian Federation.

\newpage

\begin{thebibliography}{99}

\bibitem{Kim:1997dx}
  V.~T.~Kim and G.~B.~Pivovarov,
  Phys.\ Rev.\  D {\bf 57}, 1341 (1998)
  [arXiv:hep-ph/9709433].

\bibitem{BFKL}
  L.~N.~Lipatov,
  Sov.\ J.\ Nucl.\ Phys.\  {\bf 23}, 338 (1976)
  [Yad.\ Fiz.\  {\bf 23}, 642 (1976)];
  E.~A.~Kuraev, L.~N.~Lipatov, and V.~S.~Fadin,
  Sov.\ Phys.\ JETP {\bf 44}, 443 (1976)
  [Zh.\ Eksp.\ Teor.\ Fiz.\  {\bf 71}, 840 (1976)];
  Sov.\ Phys.\ JETP {\bf 45}, 199 (1977)
  [Zh.\ Eksp.\ Teor.\ Fiz.\  {\bf 72}, 377 (1977)];
  I.~I.~Balitsky and L.~N.~Lipatov,
  Sov.\ J.\ Nucl.\ Phys.\  {\bf 28}, 822 (1978)
  [Yad.\ Fiz.\  {\bf 28}, 1597 (1978)];
  Sov.\ Phys.\ JETP {\bf 63}, 904 (1986)
  [Zh.\ Eksp.\ Teor.\ Fiz.\  {\bf 90}, 1536 (1986)].

\bibitem{Ostrovsky}
  D.~Ostrovsky,
  Phys.\ Rev.\  D {\bf 62}, 054028 (2000)
  [arXiv:hep-ph/9912258].

\bibitem{QMRK}
  V.~S.~Fadin and L.~N.~Lipatov,
  Nucl.\ Phys.\  {\bf B406}, 259 (1993);
{\bf B477}, 767 (1996)
  [arXiv:hep-ph/9602287].

\bibitem{Lipatov95}
  L.~N.~Lipatov,
  Nucl.\ Phys.\  {\bf B452}, 369 (1995)
  [arXiv:hep-ph/9502308].

\bibitem{LipatoVyazovsky}
  L.~N.~Lipatov and M.~I.~Vyazovsky,
  Nucl.\ Phys.\  {\bf B597}, 399 (2001)
  [arXiv:hep-ph/0009340].

\bibitem{SVADISy}
  V.~A.~Saleev,
  Phys.\ Rev.\  D {\bf 78}, 034033 (2008)
  [arXiv:0807.1587 [hep-ph]];
  Phys.\ Rev.\  D {\bf 78}, 114031 (2008)
  [arXiv:0812.0946 [hep-ph]].

\bibitem{SVAdiy}
  V.~A.~Saleev,
  Phys.\ Rev.\  D {\bf 80}, 114016 (2009)
  [arXiv:0911.5517 [hep-ph]].

\bibitem{PRD}
  B.~A.~Kniehl, A.~V.~Shipilova, and V.~A.~Saleev,
  Phys.\ Rev.\  D {\bf 79}, 034007 (2009)
  [arXiv:0812.3376 [hep-ph]].

\bibitem{PRb}
  B.~A.~Kniehl, V.~A.~Saleev, and A.~V.~Shipilova,
  Phys.\ Rev.\  D {\bf 81}, 094010 (2010)
  [arXiv:1003.0346 [hep-ph]];
  PoS(DIS2010), 177 (2010);
in Proceedings of Physics at the LHC 2010 (PLHC2010),
Hamburg, Germany, 2010, edited by M. Diehl, J. Haller, T. Sch\"orner-Sadenius,
and G. Steinbrueck,
DOI: \verb$http://dx.doi.org/10.3204/DESY-PROC-2010-01/shipilova$.

\bibitem{Kniehl:2006sk}
  B.~A.~Kniehl, D.~V.~Vasin, and V.~A.~Saleev,
  Phys.\ Rev.\  D {\bf 73}, 074022 (2006)
  [arXiv:hep-ph/0602179];
in Proceedings of the 15th International Workshop on Deep-Inelastic
Scattering and Related Subjects (DIS 2007),
Munich, Germany, 2007, edited by G. Grindhammer and K. Sachs,
DOI: \verb$http://dx.doi.org/10.3360/dis.2007.169$.

\bibitem{Kniehl:2006vm}
  B.~A.~Kniehl, V.~A.~Saleev, and D.~V.~Vasin,
  Phys.\ Rev.\  D {\bf 74}, 014024 (2006)
  [arXiv:hep-ph/0607254].

\bibitem{RRgCDF}
  CDF Collaboration, T.~Aaltonen {\it et al.},
  Phys.\ Rev.\  D {\bf 78}, 052006 (2008);
  {\bf 79}, 119902(E) (2009)
  [arXiv:0807.2204 [hep-ex]].

\bibitem{RRgD0}
  D0 Collaboration, V.~M.~Abazov {\it et al.},
  Phys.\ Rev.\ Lett.\  {\bf 101}, 062001 (2008)
  [arXiv:0802.2400 [hep-ex]].

\bibitem{y18}
  D0 Collaboration, B.~Abbott {\it et al.},
  Phys.\ Rev.\ Lett.\  {\bf 84}, 2786 (2000)
  [arXiv:hep-ex/9912017].

\bibitem{y196}
  D0 Collaboration, V.~M.~Abazov {\it et al.},
  Phys.\ Lett.\  B {\bf 639}, 151 (2006);
  {\bf 658}, 285(E) (2008)
  [arXiv:hep-ex/0511054].

\bibitem{Atlas}
  ATLAS Collaboration, G.~Aad {\it et al.},
  Eur.\ Phys.\ J.\  C {\bf 71}, 1512 (2011)
  [arXiv:1009.5908 [hep-ex]].

\bibitem{yAtlas}
  ATLAS Collaboration, G.~Aad {\it et al.},
  Phys.\ Rev.\  D {\bf 83}, 052005 (2011)
  [arXiv:1012.4389 [hep-ex]].

\bibitem{RRgold}
  V.~S.~Fadin and L.~N.~Lipatov,
  JETP Lett.\ {\bf 49}, 352 (1989)
 [Pis'ma Zh.\ Eksp.\ Teor.\ Fiz.\ {\bf 49}, 311 (1989)];
 Sov.\ J. Nucl.\ Phys.\ {\bf 50}, 712 (1989)
 [Yad.\ Fiz.\ {\bf 50}, 1141 (1989)];
  E.~N.~Antonov, L.~N.~Lipatov, E.~A.~Kuraev, and I.~O.~Cherednikov,
  Nucl.\ Phys.\  {\bf B721}, 111 (2005)
  [arXiv:hep-ph/0411185].

\bibitem{QR1old}
  V.~S.~Fadin and V.~E.~Sherman,
  JETP\ Lett.\ {\bf 23}, 548 (1976)
 [Pis'ma Zh.\ Eksp.\ Teor.\ Fiz.\ {\bf 23}, 599 (1976)];
 Sov.\ Phys.\ JETP {\bf 45}, 861 (1977)
 [Zh.\ Eksp.\ Teor.\ Fiz.\ {\bf 72}, 1640 (1977)].

\bibitem{KMR}
  M.~A.~Kimber, A.~D.~Martin, and M.~G.~Ryskin,
  Eur.\ Phys.\ J.\  C {\bf 12}, 655 (2000)
  [arXiv:hep-ph/9911379];
  Phys.\ Rev.\  D {\bf 63}, 114027 (2001)
  [arXiv:hep-ph/0101348];
  G.~Watt, A.~D.~Martin, and M.~G.~Ryskin,
  Eur.\ Phys.\ J.\  C {\bf 31}, 73 (2003)
  [arXiv:hep-ph/0306169];
  Phys.\ Rev.\  D {\bf 70}, 014012 (2004);
  {\bf 70}, 079902(E) (2004)
  [arXiv:hep-ph/0309096].

\bibitem{Watt} G.~Watt,
URL: \verb$http://gwatt.web.cern.ch/gwatt/$.

\bibitem{Andersson:2002cf}
  Small-$x$ Collaboration, B.~Andersson {\it et al.},
  Eur.\ Phys.\ J.\  C {\bf 25}, 77 (2002)
  [arXiv:hep-ph/0204115];
  F.~Hautmann and H.~Jung,
  Nucl.\ Phys.\ B (Proc.\ Suppl.)  {\bf 184}, 64 (2008)
  [arXiv:0712.0568 [hep-ph]].

\bibitem{Blumlein:1995eu}
  J.~Bl\"umlein,
Preprint DESY 95--121 (1995)
[arXiv:hep-ph/9506403].

\bibitem{Jung:2000hk}
  H.~Jung and G.~P.~Salam,
  Eur.\ Phys.\ J.\  C {\bf 19}, 351 (2001)
  [arXiv:hep-ph/0012143].

\bibitem{MRST}
  A.~D.~Martin, R.~G.~Roberts, W.~J.~Stirling, and R.~S.~Thorne,
  Phys.\ Lett.\  B {\bf 531}, 216 (2002)
  [arXiv:hep-ph/0201127].

\bibitem{Pumplin:2002vw}
  J.~Pumplin, D.~R.~Stump, J.~Huston, H.-L.~Lai, P.~M.~Nadolsky, and W.-K.~Tung,
  JHEP {\bf 0207}, 012 (2002)
  [arXiv:hep-ph/0201195];
  D.~Stump, J.~Huston, J.~Pumplin, W.-K.~Tung, H.-L.~Lai, S.~Kuhlmann, and
 J.~F.~Owens,
  JHEP {\bf 0310}, 046 (2003)
  [arXiv:hep-ph/0303013].

\bibitem{Gluck:1994uf}
  M.~Gl\"uck, E.~Reya, and A.~Vogt,
  Z.\ Phys.\  C {\bf 67}, 433 (1995).

\bibitem{Aversa:1988fv}
  F.~Aversa, P.~Chiappetta, M.~Greco, and J.~Ph.~Guillet,
  Phys.\ Lett.\  B {\bf 210}, 225 (1988);
{\bf 211}, 465 (1988);
  Nucl.\ Phys.\   {\bf B327}, 105 (1989);
  Z.\ Phys.\  C {\bf 46}, 253 (1990);
  Phys.\ Rev.\ Lett.\  {\bf 65}, 401 (1990);
  S.~D.~Ellis, Z.~Kunszt, and D.~E.~Soper,
  Phys.\ Rev.\ Lett.\  {\bf 62}, 726 (1989);
  Phys.\ Rev.\  D {\bf 40}, 2188 (1989);
  Phys.\ Rev.\ Lett.\  {\bf 64}, 2121 (1990);
  N.~Kidonakis,
  Int.\ J.\ Mod.\ Phys.\  A {\bf 15}, 1245 (2000)
  [arXiv:hep-ph/9902484];
  A.~Banfi, G.~P.~Salam, and G.~Zanderighi,
  JHEP {\bf 1006}, 038 (2010)
  [arXiv:1001.4082 [hep-ph]].

\bibitem{Aurenche:1987fs}
  P.~Aurenche, R.~Baier, M.~Fontannaz, and D.~Schiff,
  Nucl.\ Phys.\  B {\bf 297}, 661 (1988);
  L.~E.~Gordon and W.~Vogelsang,
  Phys.\ Rev.\  D {\bf 50}, 1901 (1994);
  J.~Huston, E.~Kovacs, S.~Kuhlmann, H.~L.~Lai, J.~F.~Owens, and W.~K.~Tung,
  Phys.\ Rev.\  D {\bf 51}, 6139 (1995)
  [arXiv:hep-ph/9501230];
  P.~Aurenche, M.~Fontannaz, J.~Ph.~Guillet, B.~Kniehl, E.~Pilon, and M.~Werlen,
  Eur.\ Phys.\ J.\  C {\bf 9}, 107 (1999)
  [arXiv:hep-ph/9811382];
  S.~Catani, M.~L.~Mangano, P.~Nason, C.~Oleari, and W.~Vogelsang,
  JHEP {\bf 9903}, 025 (1999)
  [arXiv:hep-ph/9903436];
  N.~Kidonakis and J.~F.~Owens,
  Phys.\ Rev.\  D {\bf 61}, 094004 (2000)
  [arXiv:hep-ph/9912388];
  P.~Bolzoni, S.~Forte, and G.~Ridolfi,
  Nucl.\ Phys.\  {\bf B731}, 85 (2005)
  [arXiv:hep-ph/0504115];
  D.~de Florian and W.~Vogelsang,
  Phys.\ Rev.\  D {\bf 72}, 014014 (2005)
  [arXiv:hep-ph/0506150];
  G.~Diana,
  Nucl.\ Phys.\  {\bf B824}, 154 (2010)
  [arXiv:0906.4159 [hep-ph]];
  G.~Diana, J.~Rojo, and R.~D.~Ball,
  Phys.\ Lett.\  B {\bf 693}, 430 (2010)
  [arXiv:1006.4250 [hep-ph]].

\end{thebibliography}
\end{document}